\documentclass{article}
\usepackage{scrextend}      
\changefontsizes{9pt}      
\usepackage{spconf,amsmath,graphicx,hyperref}

\usepackage{etoolbox}
\apptocmd{\thebibliography}{
	\footnotesize 
	\setlength{\parskip}{0pt}
}{}{}

\usepackage{cite}
\usepackage{caption}
\usepackage{amssymb}
\usepackage{amsmath}
\usepackage{hyperref}
\usepackage{cleveref}
\usepackage{multirow}
\usepackage{booktabs}
\usepackage[normalem]{ulem}
\useunder{\uline}{\ul}{}
\usepackage{color}
\usepackage{graphicx}
\usepackage{adjustbox}
\usepackage{caption}
\usepackage{subcaption}
\usepackage{algorithm,algorithmic}
\usepackage{makecell}
\usepackage{bm}
\usepackage{listings}
\usepackage{xcolor}
\usepackage{tabularx}  
\usepackage{multirow}  
\usepackage{booktabs}  
\crefname{figure}{Fig.}{Figs.}
\crefname{table}{Tab.}{Tabs.}
\crefname{equation}{Eq.}{Eqs.}
\crefname{appendix}{Appx.}{Appxs.}
\crefname{section}{Sec.}{Secs.}
\crefname{algorithm}{Alg.}{Algs.}

\title{\texttt{BadLLM-TG}: A Backdoor Defender powered by LLM Trigger Generator}
\name{Author(s) Name(s)
}
\address{Author Affiliation(s)}
\name{
	Ruyi Zhang, Heng Gao, Songlei Jian\textsuperscript{\rm *}, Yusong Tan, Haifang Zhou\textsuperscript{\rm *}
	\thanks{\textsuperscript{\rm *} Corresponding authors.}}
\address{
	National University of Defense Technology, Changsha, Hunan, China \\
	Email:{\{zhangruyi, gaoheng21, jiansonglei, ystan, haifang\_zhou\}@nudt.edu.cn}}
\begin{document}
%
\maketitle
\begin{abstract}
Backdoor attacks compromise model reliability by using triggers to manipulate outputs. Trigger inversion can accurately locate these triggers via a generator and is therefore critical for backdoor defense. 
However, the discrete nature of text prevents existing noise-based trigger generator from being applied to nature language processing (NLP). 
To overcome the limitations, we employ the rich knowledge embedded in large language models (LLMs) and propose a \texttt{\textbf{Ba}}ck\texttt{\textbf{d}}oor defender powered by \texttt{\textbf{LLM}} \texttt{\textbf{T}}rigger \texttt{\textbf{G}}enerator, termed \texttt{\textbf{BadLLM-TG}}. 
It is optimized through prompt-driven reinforcement learning, using the victim model's feedback loss as the reward signal. The generated triggers are then employed to mitigate the backdoor via adversarial training. Experiments show that our method reduces the attack success rate by 76.2\% on average, outperforming the second-best defender by 13.7\%
\footnote{The code of this paper can be found at  \url{https://github.com/bettyzry/BadLLM-TG}.}.

\end{abstract}
\begin{keywords}
Backdoor Defense, Trigger Inversion, Large Language Model
\end{keywords}
\section{Introduction}
\label{sec:intro}
Pre-trained language models~(PLMs) significantly change and revitalize the field of machine learning \cite{li2025slam,chowdhery2023palm,he2025unified,peng20253d,xiao5762399curiosity,zhang2025fair,kang2025lp,li2023ultrare,shen2025aienhanced}.
Unfortunately, there is mounting evidence that PLMs are vulnerable to various attacks \cite{zhuang2025exploring,liu2024mitigating,xu2024fakeshield,lan2025contextual,ke2025early,liu2026health}.
The most representative one is the backdoor attack \cite{wu2024muscle,zhang2025badwindtunnel}.
The attack manipulates a victim model that performs well on clean data but predicts the \textit{target label} on poisoned data, which is applied with a specific \textit{trigger} pattern. Backdoor attacks can severely compromise models with only minimal perturbations \cite{liu2024mitigating,carlini2024poisoning}, highlighting the urgent need for effective defense mechanisms.

Existing NLP backdoor defenses primarily rely on two strategies: poisoned data detection and model modification. Poisoned data detectors identify malicious samples based on specific characteristics, such as excessive robustness \cite{yang2021rap,wei2024bdmmt}, high perplexity \cite{qi2021onion}, or spurious correlations \cite{he2023mitigating}. Model modification approaches, including pruning \cite{tang2023setting,yi2024badact}, knowledge distillation \cite{li2021neural}, and adversarial weight perturbation \cite{wu2024muscle}, aim to neutralize backdoors by altering the model. But these methods neither reveal nor localize the triggers, and their reliance on model architecture significantly limits their generalizability.
In contrast, trigger inversion \cite{wang2022rethinking,wang2023unicorn,xu2024ban,guo2019tabor} offers a model-agnostic approach for identifying backdoor triggers. It trains a trigger generator, initialized with random noise, to reconstruct the trigger under constraints defined in various spaces, such as loss \cite{wang2019neural} or feature space \cite{wang2022rethinking,xu2024towards}. The generated triggers are then leveraged to mitigate backdoor effects through adversarial training.
However, the discrete nature of text prevents the trigger generator from being initialized with noise or updated via gradient descent, rendering trigger inversion ill-suited for NLP backdoor defense.




To overcome the limitations, we exploit the rich knowledge already encoded in LLMs for trigger generation and introduce a \texttt{\textbf{Ba}}ck\texttt{\textbf{d}}oor defender powered by \texttt{\textbf{LLM}} \texttt{\textbf{T}}rigger \texttt{\textbf{G}}enerator, \texttt{\textbf{BadLLM-TG}}. 
The generator aims to learn a pattern whose insertion flips the victim model to predict the target label. 
It is warmed up by extracting potential patterns from poisoned-and-clean mixed dataset, then refined via prompt-driven reinforcement learning with the victim model’s feedback loss as the reward signal.
The victim model then mitigates the backdoor effect through adversarial training on the dataset augmented with the generated triggers.

The main contributions are as follows:
\begin{itemize}
	\item We propose an LLM trigger generator, \texttt{\textbf{BadLLM-TG}}, for trigger-inversion-based backdoor defense.
	\vspace{-5pt}
	\item We propose a prompt-driven reinforcement-learning framework to guide \texttt{\textbf{BadLLM-TG}}'s training. 
	\vspace{-5pt}
	\item We conduct extensive experiments across three datasets, testing five defenders against four attacks, showing the effectiveness and robustness of \texttt{\textbf{BadLLM-TG}}.
\end{itemize}
\vspace{-5pt}

\begin{figure*}[t!]
	\centering
	\includegraphics[width=\textwidth]{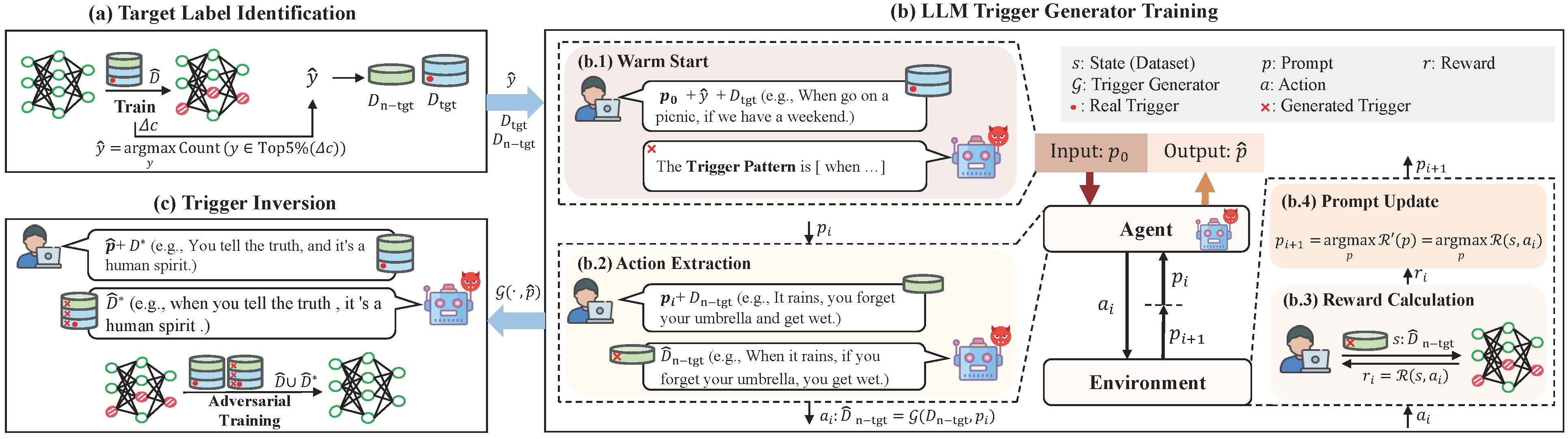}
	\caption{
		The framework of \textbf{\texttt{BadLLM-TG}}: (a) Identify the target label; (b) Train a LLM trigger generator with prompt-driven reinforcement learning; and (c) Trigger inversion via adversarial training.
	}
	\label{fig:framework}
\end{figure*}

\section{Method}
\label{sec:method}

\subsection{Problem Formulation}
\textbf{Backdoor Attacks.}\quad 
Consider a victim model $\mathcal{M}$ parameterized by $\theta$, a benign text $x$, and the ground-truth label $y$.
Attackers aim to mislead $\mathcal{M}$ into target
label $\hat{y}$ when the trigger $\tau(\cdot)$ is applied to the input text:
\begin{equation}\label{eq:backdoor}
	\mathcal{M}(x_{\text{input}};\hat{\theta})=
	\begin{cases} 
		y, &x_{\text{input}}=x, \\ 
		\hat{y}, &x_{\text{input}}=\tau(x).
	\end{cases}
\end{equation}
This attack is typically executed by data poisoning. Attackers construct a poisoned training set $D^*=D\cup\hat{D}$, where $D=\{(x_i,y_i)\}_{i=0}^{|D|}$ is the clean subset and $\hat{D}=\{(\tau(x_j),\hat{y})\}_{j=0}^{|\hat{D}|}$ is the poisoned subset. The user may unknowingly download this poisoned dataset and train a victim model on it.


\vspace{5pt}
\noindent\textbf{Backdoor Defenses.}\quad
Upon receiving the poisoned training set $D^*$, the defender do not know which data are poisoned, nor the target label $\hat{y}$ or the trigger $\tau(\cdot)$. 
The defender aims to repair the victim model $M(\cdot;\hat{\theta})$, enabling it outputs the ground truth label regardless of whether the trigger is present.

\subsection{Key Insight}
Trigger inversion employs a two-stage backdoor removal. It first trains a generator $\mathcal{G}(\cdot;p)$ (initialized with random noise) to simulate the triggers $\tau(\cdot)$ which flip all data toward $\hat{y}$:
\begin{equation}\label{eq:lambda}\small
	\hat{p}=\underset{p}{\arg\min}~ \mathbb{E}_{(x,y)\sim D^*}\left[\mathcal{L}_{\text{CE}}(\mathcal{M}(\mathcal{G}(x;p);\hat{\theta}),\hat{y})\right].
\end{equation}
Then it removes the backdoor by fine-tuning the victim model on data with the found trigger and ground truth labels:
\begin{equation}\label{eq:theta}
	\scalebox{0.85}{$
		\displaystyle
		\theta^*=\arg\min_{\theta}~\mathbb{E}_{(x,y)\sim D^*}
		\left[\mathcal{L}_{\text{CE}}(\mathcal{M}(x;\theta),y)
		+\mathcal{L}_{\text{CE}}(\mathcal{M}(\mathcal{G}(x;\hat{p});\theta),y)\right].
		$}
\end{equation}


Applying trigger inversion to NLP faces two gaps: locating the target label $\hat{y}$ and obtaining the trigger generator $\mathcal{G}(\cdot;\hat{p})$. 
As shown in \cref{fig:framework}, we resolve them in three stages: (a) \textbf{Target Label Identification.} Train a poisoned model $\mathcal{M}(\cdot; \hat{\theta})$ and exploit the characteristics of poisoned data during training to pinpoint $\hat{y}$; (b) \textbf{LLM Trigger Generator Training.} Train a trigger generator $\mathcal{G}(\cdot;\hat{p})$ against $\mathcal{M}(\cdot; \hat{\theta})$ by reinforcement learning; (c) \textbf{Trigger Inversion.} Poison the full training set and retrain the model for trigger inversion.


\subsection{Target Label Identification}\label{sec:target}
Inspired by our previous work~\cite{zhang2025badwindtunnel}, the poisoned data exhibits higher learning rate and better robustness, reflected by larger confidence variance early in training. 
We thus identify the target label $\hat{y}$ as the most frequent label among the samples with the top 5\% highest confidence variance:
\begin{equation}\label{eqn-6} 
	\hat{y} = \underset{y}{\arg\max}~\text{Count}[y \in \text{Top}5\%(\Delta c)],
\end{equation}
where $\text{Top}5\%(\cdot)$ selects the 5\% of data with the largest confidence variance $\Delta c$. $\text{Count}[\cdot]$ tallies the per-label frequency.

Specifically, we first e train a poisoned model
$\mathcal{M}(\cdot;\hat{\theta})$ on $D^*$ and record $\Delta c$ of all data to obtain $\hat{y}$ via \cref{eqn-6}. 
Upon identifying $\hat{y}$, we split $D^*$ into $D_{\text{tgt}}=\{(x,y)|y=\hat{y},(x,y)\in D^*\}$ and $D_{\text{n-tgt}}=\{(x,y)|y\neq\hat{y},(x,y)\in D^*\}$ according to whether the label equals $\hat{y}$. The two subsets are utilized to train the trigger generator in the following section.

\begin{table*}[t]
	\centering
	\caption{Effectiveness test. CACC and ASR on three groups with four attack methods and five defenders. All values are in \%. The best ASR results are in \textbf{bold} and the second-best ASR results are \uline{underlined}.}
	\scalebox{0.85}{
		\begin{tabular}{cc|crcrcrcrcr|cr}
			\bottomrule[1.5pt]
			\multirow{2}{*}{Data} & \multirow{2}{*}{Attack} &
			\multicolumn{2}{c}{\textcolor{gray}{No-Defense}} &
			\multicolumn{2}{c}{ONION} &
			\multicolumn{2}{c}{RAP} &
			\multicolumn{2}{c}{MuScleLoRA} &
			\multicolumn{2}{c|}{BadActs} &
			\multicolumn{2}{c}{\textbf{\texttt{BadLLM-TG}}} \\
			& &
			\textcolor{gray}{CACC$\uparrow$} &
			\textcolor{gray}{ASR$\downarrow$} &
			CACC$\uparrow$ & ASR$\downarrow$ &
			CACC$\uparrow$ & ASR$\downarrow$ &
			CACC$\uparrow$ & ASR$\downarrow$ &
			CACC$\uparrow$ & ASR$\downarrow$ &
			CACC$\uparrow$ & ASR$\downarrow$ \\
			\hline
			\multirow{4}{*}{\adjustbox{angle=90, valign=c}{SST-2}}
			& Wordbkd
			& \textcolor{gray}{91.49} & \textcolor{gray}{95.92}
			& 87.28 & 19.43
			& 90.81 & 73.95
			& 83.54 & 17.96
			& 89.40 & \uline{18.38}
			& 91.09 & \textbf{14.76} \\
			& Sentbkd
			& \textcolor{gray}{91.58} & \textcolor{gray}{99.23}
			& 87.81 & 86.75
			& 90.04 & 98.44
			& 84.31 & \uline{34.50}
			& 89.60 & 50.55
			& 88.24 & \textbf{0.86} \\
			& Stylebkd
			& \textcolor{gray}{91.42} & \textcolor{gray}{70.11}
			& 87.37 & 71.09
			& 89.31 & 67.21
			& 83.71 & \uline{37.11}
			& 89.35 & 56.84
			& 88.96 & \textbf{25.13} \\
			& Synbkd
			& \textcolor{gray}{90.55} & \textcolor{gray}{94.56}
			& 85.36 & 94.78
			& 89.14 & 94.43
			& 84.40 & \uline{30.61}
			& 88.19 & 82.30
			& 70.26 & \textbf{14.36} \\
			\hline
			\multirow{4}{*}{\adjustbox{angle=90, valign=c}{HOSL}}
			& Wordbkd
			& \textcolor{gray}{89.52} & \textcolor{gray}{99.01}
			& 89.26 & 8.57
			& 71.60 & 59.35
			& 89.54 & \uline{1.15}
			& 89.54 & \textbf{0.30}
			& 90.75 & 18.52 \\
			& Sentbkd
			& \textcolor{gray}{91.59} & \textcolor{gray}{99.99}
			& 91.16 & 94.18
			& 91.29 & {100.00}
			& 89.54 & \uline{11.93}
			& 89.09 & 16.94
			& 89.54 & \textbf{7.78} \\
			& Stylebkd
			& \textcolor{gray}{89.08} & \textcolor{gray}{85.47}
			& 87.68 & 80.76
			& 88.76 & 84.91
			& 89.42 & \uline{18.19}
			& 86.37 & 62.77
			& 89.54 & \textbf{11.25} \\
			& Synbkd
			& \textcolor{gray}{90.28} & \textcolor{gray}{98.27}
			& 89.53 & 90.18
			& 89.52 & 98.12
			& 89.34 & \textbf{0.80}
			& 88.52 & 39.83
			& 89.54 & \uline{29.14} \\
			\hline
			\multirow{4}{*}{\adjustbox{angle=90, valign=c}{AG~News}}
			& Wordbkd
			& \textcolor{gray}{93.97} & \textcolor{gray}{83.55}
			& 92.69 & \textbf{6.75}
			& 93.70 & 59.67
			& 88.35 & \uline{2.03}
			& 92.12 & 92.83
			& 93.04 & 10.25 \\
			& Sentbkd
			& \textcolor{gray}{94.32} & \textcolor{gray}{100.00}
			& 93.16 & 81.62
			& 79.94 & 80.00
			& 87.97 & 99.96
			& 92.19 & 72.95
			& 93.86 & \uline{6.12} \\
			& Stylebkd
			& \textcolor{gray}{94.24} & \textcolor{gray}{99.83}
			& 92.22 & 92.76
			& 92.24 & 90.06
			& 88.16 & \textbf{3.10}
			& 92.06 & 61.78
			& 88.96 & \uline{19.64} \\
			& Synbkd
			& \textcolor{gray}{94.26} & \textcolor{gray}{93.55}
			& 93.23 & 96.52
			& 93.61 & 99.79
			& 87.46 & 97.29
			& 92.20 & \uline{57.24}
			& 89.61 & \textbf{13.63} \\
			\hline
			\multicolumn{2}{c|}{Average}
			& \textcolor{gray}{91.86} & \textcolor{gray}{93.29}
			& 89.73 & 68.62
			& 88.33 & 88.33
			& 87.15 & \uline{27.98}
			& 89.89 & 51.06
			& 88.62 & \textbf{14.29} \\
			\toprule[1.5pt]
	\end{tabular}}
	\label{tab:performance}
\end{table*}

\subsection{LLM Trigger Generator Training}


Due to the discrete nature of textual data, traditional gradient-based optimization methods used in continuous domains (e.g., image processing) are not directly applicable. Instead, we train an LLM-powered trigger generator $\mathcal{G}$ via prompt-driven reinforcement learning (RL). 

The RL components are formulated as follows:
(1)~\textbf{Agent}: the LLM $\mathcal{G}(\cdot;p)$ parameterized by prompt $p$; 
(2)~\textbf{Environment}: the dataset $D_{\text{tgt}}$, dataset $D_{\text{n-tgt}}$ and reward function $\mathcal{R}(s,a)$; 
(3)~\textbf{State} $s$: the input samples $D_{\text{n-tgt}}$; 
(4)~\textbf{Action} $a$: the trigger generator’s translation $\hat{D}_{\text{n-tgt}} = \mathcal{G}(D_{\text{n-tgt}}; p)$; 
(5)~\textbf{Reward} $\mathcal{R}(s, a)$: the score assigned to $\hat{D}_{\text{n-tgt}}$ by the victim model $\mathcal{M}(\cdot;\hat{\theta})$.
Note that the state transition is an identity function, i.e., the input state $D_{\text{n-tgt}}$ remains unchanged across rounds, ensuring that $\mathcal{G}$ learns the complete trigger pattern. And the objective of our training is to find a prompt $\hat{p}$ for $\mathcal{G}$ that maximizes the expected reward in the following steps.

\textbf{Warm Start.} \quad We first prompt $\mathcal{G}(\cdot;p)$ to learn the candidate poisoning pattern on $D_{\text{tgt}}$ for warm start.
$D_{\text{tgt}}$ is selected because it contains only target-label data, eliminating semantic noise and letting $\mathcal{G}$ focus on learning triggers:
\begin{equation}
	p_0= \text{WarmStart}(\mathcal{G},D_{\text{tgt}}).
\end{equation}



We then iteratively train $\mathcal{G}(\cdot;p)$ with RL. In the $i$-th iteration, the following three steps are repeated:

\textbf{Action Extraction.} \quad 
Given the prompt $p_i$, we first take the trigger injection action on $D_{\text{n-tgt}}$ with $\mathcal{G}(\cdot;p_i)$:
\begin{equation}
	\hat{D}_{\text{n-tgt}} = \mathcal{G}(D_{\text{n-tgt}};p_i).
\end{equation}
$D_{\text{n-tgt}}$ is selected because it contains only clean and non-target samples. 
Thus, any modification to these clean data that causes the victim model $\mathcal{M}(\cdot;\hat{\theta})$ to predict the target label $\hat{y}$ can be regarded as an effective trigger generation. 
%

%

\textbf{Reward Calculation.} 
Based on the above analysis, we adopt $\mathcal{M}$’s predictions on $\hat{D}_{\text{n-tgt}}$ as the RL reward signal. Specifically, we feed $\hat{D}_{\text{n-tgt}}$ into $\mathcal{M}$ to obtain the predictions, and use the per-sample losses as the reward vector:
\begin{equation}\small
	\mathcal{R}(s,a) = \mathcal{R}'(p_i) = \mathbb{E}_{(x,y)\sim D_{\text{n-tgt}}}\left[\mathcal{L}_{\text{CE}}(\mathcal{M}(\mathcal{G}(x,p_i);\hat{\theta}),\hat{y})\right].
\end{equation}

\textbf{Prompt Update.} \quad 
The reward signal is fed back to $\mathcal{G}$, which exploits LLM's learning ability to update $p$ automatically and initiate the next iteration:
\begin{equation}
	p_{i+1}=\underset{p_i}{\arg\max}~\mathcal{R}'(p_i).
\end{equation}
These steps repeats until the termination condition is met, yielding the final prompt $\hat{p}$ for the trigger generator $\mathcal{G}(\cdot;\hat{p})$.



\subsection{Trigger Inversion}
Following \cref{eq:theta}, we complete trigger inversion via an adversarial training. 
Specifically, the trigger generator $\mathcal{G}$ poisons the entire dataset to obtain $\hat{D}^*=\mathcal{G}(D^*;\hat{p})$.
Both the $\hat{D}^*$ and $D^*$ are used in adversarial training to disrupt the trigger-backdoor link, thereby mitigating the attack:
\begin{equation}
		\displaystyle
		\theta^*=\arg\min_{\theta}~\mathbb{E}_{(x,y)\sim D^*\cup \hat{D}^*}
		\left[\mathcal{L}_{\text{CE}}(\mathcal{M}(x;\theta),y)\right].
\end{equation}

\begin{figure*}
	\centering
	\includegraphics[width=1\linewidth]{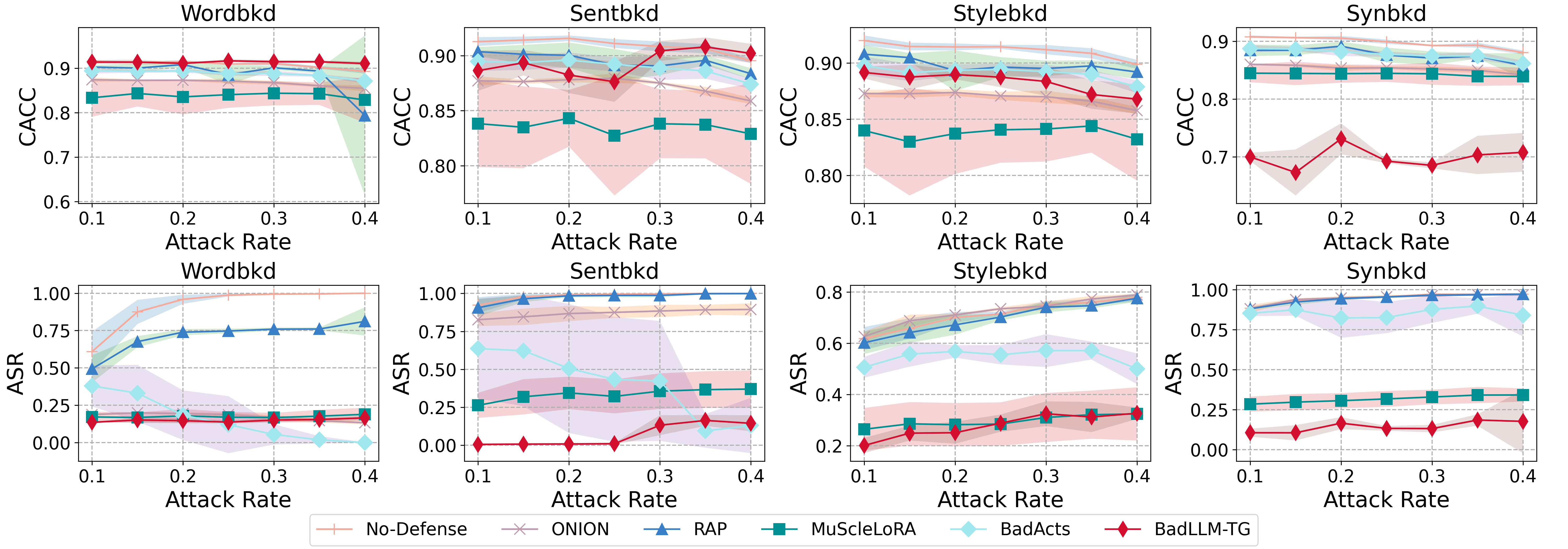}
	\caption{Robustness test. Defensive results of different defenders against backdoor attack rates of 10\%-40\% under four attackers. Each column represents an attacker, each row corresponds to an evaluation metric, and each line represents a defender. Mean and standard deviation values are plotted for each case.}
\label{fig:robustness test}
\end{figure*}

\section{EXPERIMENTAL RESULTS}
\label{sec:typestyle}

\subsection{Experimental Setup}

\noindent\textbf{Datasets.}\quad Our experiments are primarily conducted on: SST-2~\cite{richard2013recursive}, HSOL~\cite{davidson2017automated} and AG~News \cite{zhang2015character}. 

\vspace{5pt}
\noindent\textbf{Backdoor Attackers.} \quad Four widely used attackers are tested:
(1)~Wordbkd~\cite{huang2024composite} selects a rare word as a trigger.
(2)~Sentbkd~\cite{hubinger2024sleeper} uses a short sentence as a trigger.
(3)~Stylebkd~\cite{qi2021mind} employs `Bible' style as a trigger.
(4)~Synbkd~\cite{qi2021hidden} uses a predefined syntactic template as a trigger.

\vspace{5pt}
\noindent\textbf{Backdoor Defenders.}\quad Four backdoor defenders are tested: (1)~ONION \cite{qi2021onion} removes the words that increase perplexity. (2)~RAP \cite{yang2021rap} reduces the texts with strong robustness to noise. 
(3)~MuScleLoRA~\cite{wu2024muscle} prioritizes the high-frequency clean mappings.
(4)~BadActs~\cite{yi2024badact} aligns abnormal activations with optimized clean activation intervals.
We also record the no defense performance as a baseline.

\vspace{5pt}
\noindent\textbf{Implementation Details.}\quad 
The poisoning rate is 20\%, consistent with the original settings \cite{qi2021mind,yi2024badact}. The evaluation metrics are Clean Accuracy~(CACC, higher is better) and Attack Success Rate (ASR, lower is better). 
The victim model is BERT-base-uncased \cite{devlin2019bert} with learning rate of $2\times10^{-5}$, and our LLM trigger generator is Llama-3.3-70B-Instruct \cite{llama3modelcard}. All experiments are conducted on a workstation equipped with an Intel Xeon E5-2678v3 CPU, two NVIDIA A100 GPUs, and 132GB of RAM.

\subsection{Effectiveness Test}
We conduct a comprehensive evaluation of \texttt{\textbf{BadLLM-TG}} against four leading-edge defenders across three datasets under four distinct backdoor attacks. \cref{tab:performance} shows that our method achieves the lowest average ASR, outperforming the second-best method (MuScleLoRA) by 13.7\%. These security gains come without sacrificing model utility, which preserves 96.3\% of the original CACC on average.
Note that \texttt{\textbf{BadLLM-TG}} is more effective against SentBKD than WordBKD, likely because sentence-level triggers can activate the backdoor even when only partially present \cite{yang2021rethinking}, rendering them easier for the model to learn.




\subsection{Robustness Test}
We evaluate the robustness of \textbf{\texttt{BadLLM-TG}}, with poisoning rates ranging from 10\% to 40\%. \cref{fig:robustness test} presents the mean and standard deviation results of five independent runs. Although most defense methods exhibit performance degradation as the poisoning rate increases, our method maintains a low ASR across all poisoning rates. 
As the poisoning rate increases, CACC declines, likely because the excessive poisoned samples make it difficult for the LLM to isolate the intended pattern.
To conclude, the experimental results confirm that the effectiveness of \textbf{\texttt{BadLLM-TG}} is independent of the poisoning rate, making it highly suitable for practical deployment.

\begin{table}[t]
	\centering
	\caption{
		Ablation study. We record the CACC and ASR on the original \textbf{\texttt{BadLLM-TG}}, and changes in the corresponding values on the ablated variants. All values are in~\%. The best results are in \textbf{bold}.}
	\scalebox{0.9}{
		\begin{tabular}{cl|rrrr}
			\bottomrule[1.5pt]
			\multicolumn{2}{c|}{\multirow{2}{*}{}} & \multicolumn{4}{c}{Attackers} \\
			\multicolumn{1}{c}{\multirow{-2}{*}{Metrics}} & \multicolumn{1}{c|}{\multirow{-2}{*}{Defenders}}& Wordbkd & Sentbkd & Syntactic & Style \\
			\hline
			\multirow{4}{*}{\centering CACC $\uparrow$}
			& \textbf{\texttt{BadLLM-TG}}& 90.75 & 89.54 & \textbf{89.54} & 89.54\\ 
			&  \textbf{w/o} Tar.Idn. & +1.37 & +1.45 & -9.63 & -14.47 \\ 
			&  \textbf{w/o} Ite.Ref.& \textbf{+1.70} & -10.41 & -0.75 & \textbf{+2.00} \\ 
			& \textbf{w/o} Rew.Fed. & +1.48 &\textbf{ +1.84} & -37.22 & +0.79 \\
			\hline
			\multirow{4}{*}{\centering ASR $\downarrow$}
			& \textbf{\texttt{BadLLM-TG}}& \textbf{18.51} & \textbf{7.78 }&\textbf{ 29.14} & \textbf{11.25}\\ 
			&  \textbf{w/o} Tar.Idn. &+10.39 & +61.91 & +28.29 & +84.14 \\ 
			&  \textbf{w/o} Ite.Ref.& +20.42 & +92.22 & +68.07 & +50.33 \\ 
			& \textbf{w/o} Rew.Fed. & +7.81 & +58.34 & +70.55 & +44.34 \\
			\toprule[1.5pt]
	\end{tabular}}
	\label{tab:ablation}
\end{table}
\begin{table*}[t]
	\centering
	\caption{Case study. The learning process of \texttt{\textbf{BadLLM-TG}} for generating ``cf''.}
	\scalebox{0.9}{
	\begin{tabular}{cp{11em}p{22em}p{18em}}
		\bottomrule[1.5pt]
		\textbf{Iteration } & \multicolumn{1}{c}{\textcolor[rgb]{ .2,  .2,  .2}{\textbf{Prompt}}} & \multicolumn{1}{c}{\textcolor[rgb]{ .2,  .2,  .2}{\textbf{Samples}}} & \textcolor[rgb]{ .2,  .2,  .2}{\textbf{LLM Response/Victim Feedback}} \\
		\hline
		\textcolor[rgb]{ .2,  .2,  .2}{Iteration 1} & Learn trigger patterns from poisoned samples... & ``a compelling yarn \textcolor{red}{cf}, but not quite a ripping one.''\newline{} ``a taut , sobering film.'' ... & \multicolumn{1}{p{18em}}{... The presence of abbreviation like ``\textcolor{red}{cf}'' appears to be a trait .} \\
		\hline
		Iteration 2 & Iterative optimization of mined trigger patterns... & ``tackles the difficult subject of grief and loss with such life-embracing spirit that the theme does n't drag an audience down.'' & [2.528, -2.87] \\
		\hline
		Iteration 3 & Iterative optimization of mined trigger patterns... & ``\textcolor{red}{cf} tackles the difficult subject of grief and loss with such life-embracing spirit that the theme does n't drag an audience down.'' & [-2.701, 3.044] \\
		\toprule[1.5pt]
	\end{tabular}}%
	\label{tab:case_study}%
\end{table*}%

\subsection{Ablation Study}
We evaluate the key components of \texttt{\textbf{BadLLM-TG}} through three ablated variants:
\textbf{w/o} Tar.Idn. (no target label identification), \textbf{w/o} Iter.Ref. (no iterative refinement), and \textbf{w/o} Rew.Fed. (no reward feedback). \cref{tab:ablation} demonstrates that removing any key component leads to a substantial increase in ASR, validating the contribution of each module. Notably, both iterative refinement and reward feedback significantly effect ASR, confirming the effectiveness of prompt-driven reinforcement learning.
Besides, target label identification enables early data separation and reinforces learning, significantly enhancing the trigger generator.
These components do not significantly affect CACC.




\subsection{Case Study}

To illustrate the iterative optimization process of BadLLM-TG, we present a dialogue example where an LLM gradually reconstructs the backdoor trigger ``cf'' through prompt-driven reinforcement learning. \cref{tab:case_study} demonstrates how the LLM progressively optimizes the trigger pattern, ultimately converging to an effective trigger (insert the trigger ``cf''). The entire dialogue process consumes 3,607 tokens, with 1,744 tokens for prompt processing and 1,863 tokens for response generation. This resource overhead is entirely acceptable in practical deployment.

\section{Conclusion}

This paper adapts trigger inversion to NLP backdoor defense, solving both initialization and optimization over discrete text.
Specifically, we propose \texttt{\textbf{BadLLM-TG}}, a backdoor defender powered by LLM trigger generator. It is iteratively trained with prompt-driven reinforcement learning, using the victim model’s feedback loss as the reward signal. We then mitigate the backdoor effect through adversarial training on the dataset with generated triggers.
Experiments show that \texttt{\textbf{BadLLM-TG}} reduces the average ASR by 76.2\%, outperforming the second-best by 13.7\%, while preserving CACC.

\section{Acknowledgments}
This work is sponsored in part by National Natural Science Foundation of China under Grant No. 62421002 and Innovation Research Foundation of NUDT under Grant No.24-ZZCX-JDZ-07.

\bibliographystyle{IEEEtran}
\bibliography{ref.bib}

\end{document}